\documentclass[
    ,final            
  ,numberedheadings 
  ]
  {aipproc}

\layoutstyle{8x11double}

\usepackage{amssymb,amsmath,caption,color}

\usepackage{natbib}

\begin{document}

\newcommand{\daniil}[1]{{\color{blue} #1}}
\newcommand{\michal}[1]{{\color{green} #1}}
\newcommand{\mathieu}[1]{{\color{red} #1}}

\renewcommand{\floatpagefraction}{1.0}
\renewcommand{\textfraction}{0.}

\newcommand{\etal}{{\it{et al }}}
\newcommand{\apj}{ApJ}
\newcommand{\aap}{AAP}
\newcommand{\nat}{Nature}
\newcommand{\mnras}{{\em Mon. Not. R. Ast. Soc.}}

\title{Morphology and hardness ratio exploitation under limited statistics}



\author{Clapson A.-C.}{
  address={MPIK}
}

\author{Dyrda M.}{
  address={ INP PAN, Cracow, Poland}  
  ,altaddress={MPIK} 
}

\author{Nekrassov D.}{
  address={MPIK}
}

\author{Renaud M.}{
  address={MPIK}
}

\begin{abstract}

$\gamma$-ray astronomy has produced for several years now sky maps
for low photon statistics, non-negligible background and comparatively poor angular resolution.
Quantifying the significance of spatial features remains difficult. 
Besides, spectrum extraction requires regions with large statistics 
while maps in energy bands allow only qualitative interpretation.\\
The two main competing mechanisms in the VHE domain are the 
Inverse-Compton emission from accelerated electrons radiating 
through synchrotron in the X-ray domain and the interactions 
between accelerated hadrons and the surrounding medium, leading 
to the production and subsequent decay of $\pi^{0}$ mesons. 
The spectrum of the VHE emission from leptons is predicted to steepen with 
increasing distance from the acceleration zone, owing to 
synchrotron losses (i.e. cooled population). It would remain 
approximately constant for hadrons.

Ideally, spectro-imaging analysis would have the same spatial 
scale in the TeV and X-ray domains, to distinguish the local emission mechanisms. 
More realistically, we investigate here the possibility of improving upon the currently 
published HESS results by using more sophisticated tools.

\end{abstract}

\maketitle


\section{Description of methods}
\begin{figure}[h!]
\begin{minipage}[c][4.6cm][c]{0.3\textwidth}
\centering
\includegraphics[height=4.5cm,angle=0]{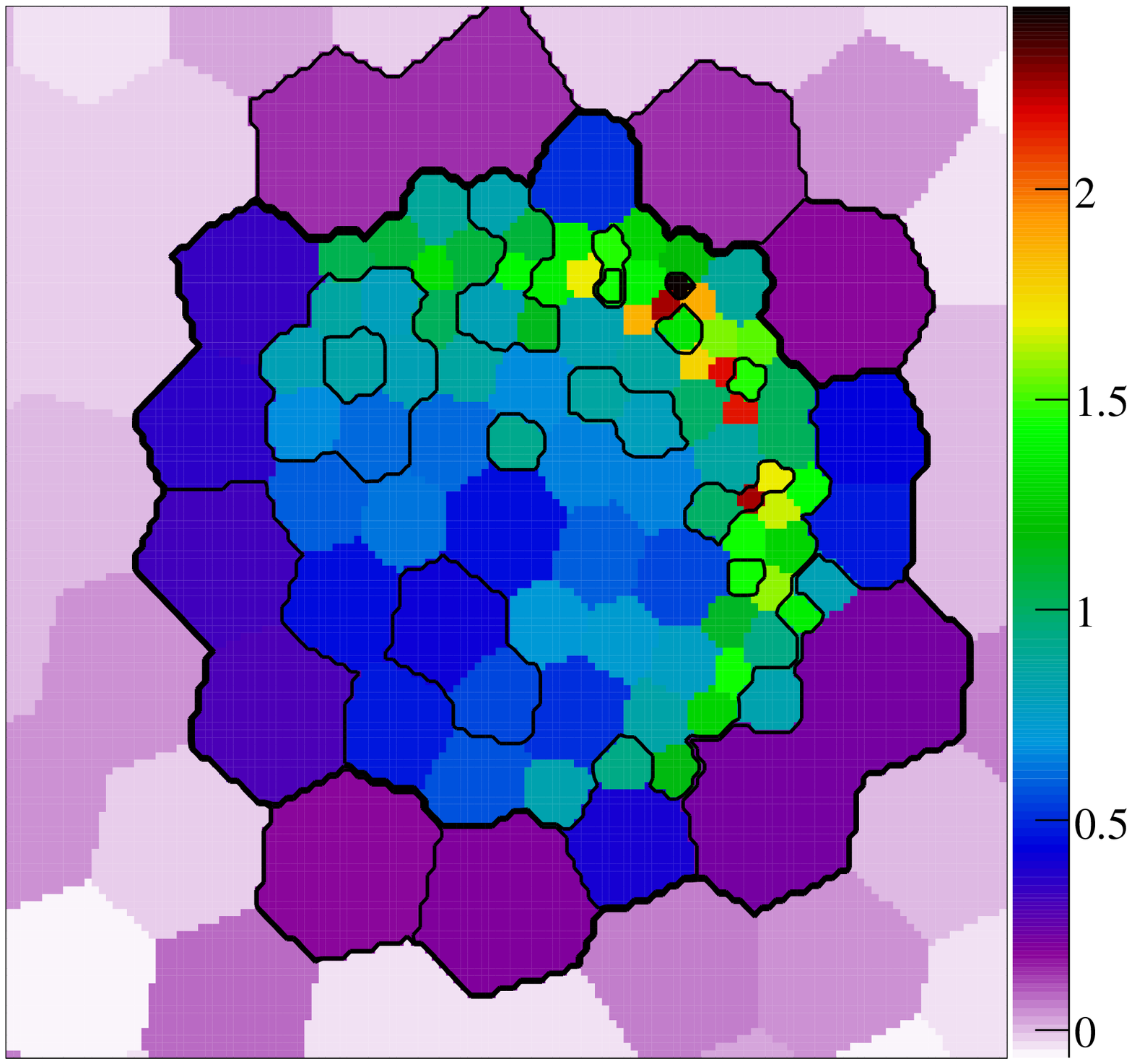}
\end{minipage}
\begin{minipage}[c][4.6cm][c]{0.3\textwidth}
\centering
\includegraphics[height=4.5cm,angle=0]{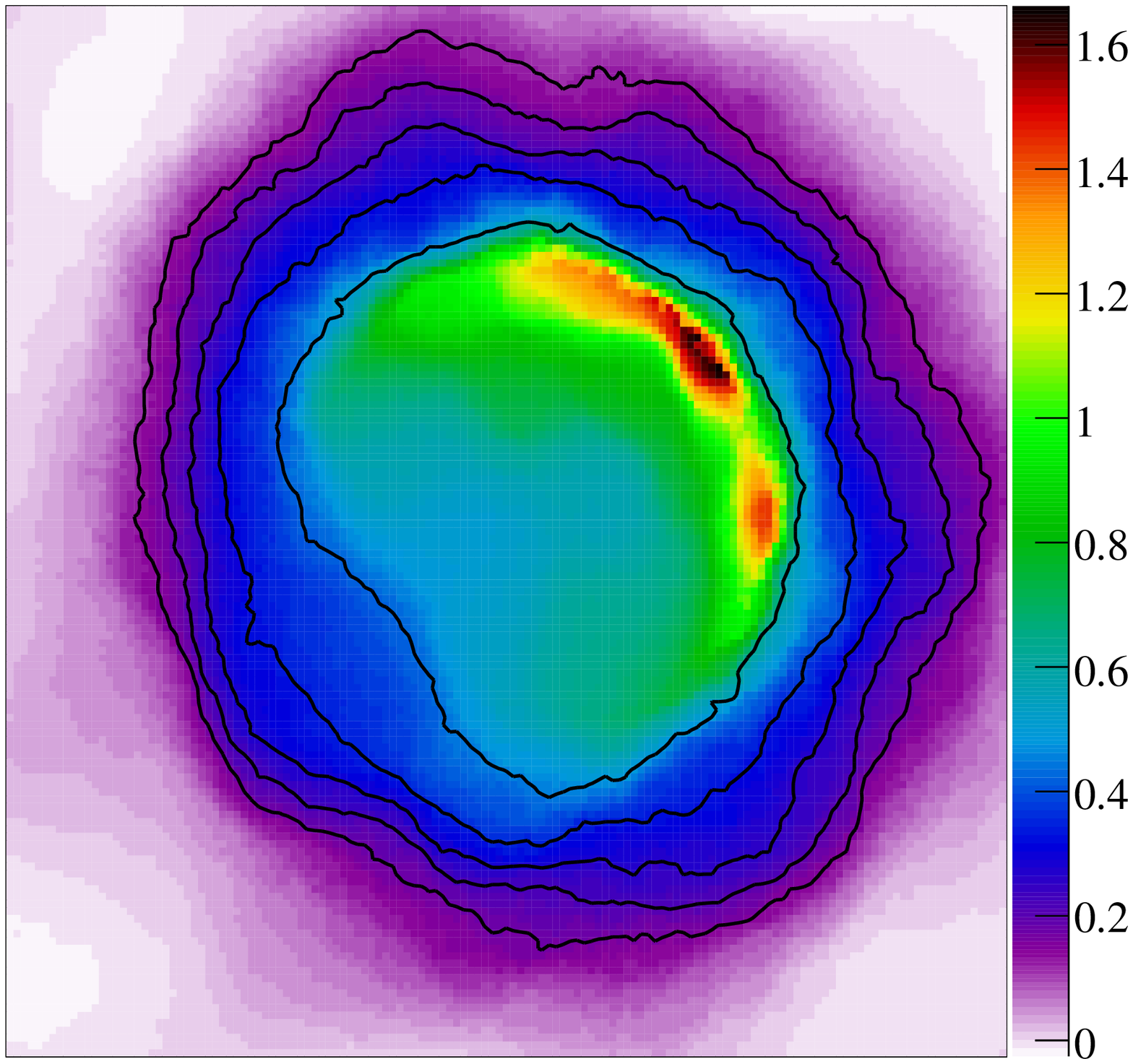}
\end{minipage}
\begin{minipage}[c][4.6cm][c]{0.3\textwidth}
\centering
\includegraphics[height=4.5cm,angle=0]{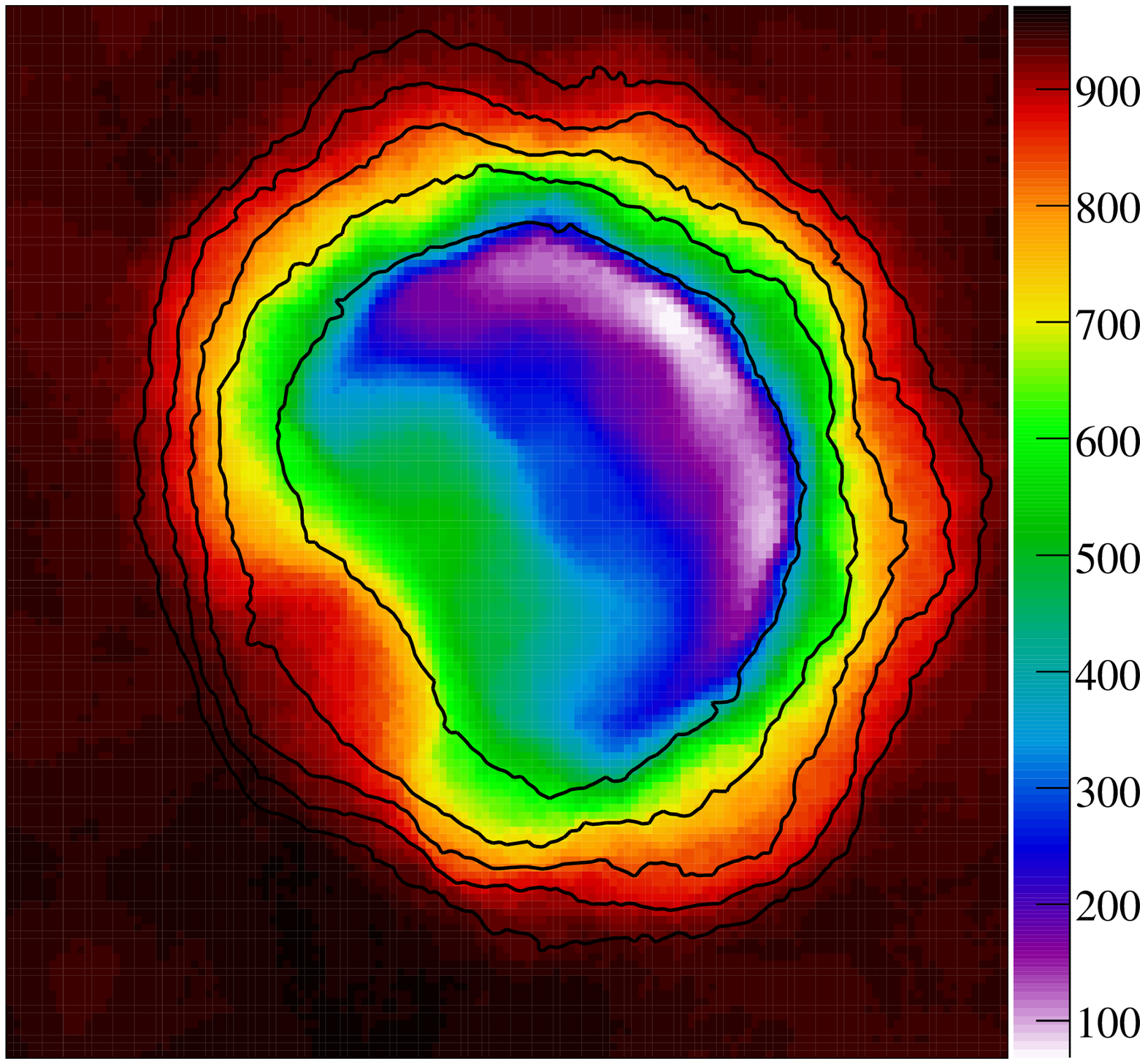}
\end{minipage}
\caption{The WVT algorithm. From {\it left} to {\it right}, the original WVT, the adapted WVT 
and the associated average bin sizes. Significance contours from 3 to 7 $\sigma$.
\label{figure:methods2} }
\end{figure}

The most simple procedure to produce hardness maps relies on bins of
fixed size, defined independently of the source morphology.
The bin size is a compromise between increasing the number of
excess events per bin and keeping small-scale structures.
This guarantees the absence of any internal artefact but 
is not adapted in case of large flux variations.
\newline
{\bf Core equations}\\
\begin{eqnarray*}
E \, &= &N_{\mathrm{on}} - \frac{Exposure_{on}}{Exposure_{off}} \times N_{\mathrm{off}} \quad
\mathrm{C}=\frac{N_{\mathrm{high}} -
  N_{\mathrm{low}}}{N_{\mathrm{high}} + N_{\mathrm{low}}} \\
  S_C &= &\frac{S_{low} \times S_{high}}{\sqrt{S_{low}^2 + S_{high}^2}}
\end{eqnarray*}
\noindent For $\gamma$-like ($\mathrm{on}$) and background ($\mathrm{off}$) events
in two separate high and low energy bands, these formula define the excess $E$~\cite{berge},
the contrast $C$ (used here instead of the immediate ratio of excesses) 
between the bands and its significance $S_C$,
combining the significance in each band, given by the Li\&Ma formula~\cite{lima}.
The selection of the energy bands, starting at the energy threshold
for the observation, aims at balanced counts of excess events.
\newline
{\bf Adaptive binning}\\
In the Weighted Voronoi Tessellation (WVT) adaptive binning algorithm
\cite{wvtpaper}, the core procedure assigns pixels to
non-overlapping bins, whose shape is maintained close to circular.
From a given position $P_{start}$, neighbouring pixels are accreted
until the requested significance or maximum size is reached. 
The assignment of pixels is repeated to reduce
the spread of bin significance and shapes.
Identified point sources are treated here as fixed bins, to avoid contamination 
by the background as well as smearing out of the source.
To make the whole algorithm independent of the start position,
we loop over random $P_{start}$, keeping the average of the re-binned maps.
This procedure is illustrated in Fig.~\ref{figure:methods2}.
The target significance is 7 $\sigma$,
with an upper limit on bin radius of 0.12$^{\circ}$.
\newline
{\bf Adaptive smoothing}\\
Adaptive smoothing methods, like ASMOOTH~\cite{asmooth} algorithm (CSMOOTH in the CIAO package
\cite{ciao}), adjust the scale of a smoothing kernel to maintain a constant significance.
For a given scale, the image is smoothed and pixels reaching the desired 
significance are added to the output image and removed from the input one. 
Remaining pixels are treated with progressively larger scales.
To build the contrast map, the two energy bands are treated separately, see~\cite{asmooth} for details,
with a 9 $\sigma$ target here.
\newline
{\bf Photon index}\\
From the contrast, we estimate the photon index and flux 
normalization, assuming a power-law spectrum.
A hardness ratio derived from excess counts instead of fluxes
is affected by the instrument response. To correct for this,
the effective area of the instrument is integrated in the energy bands, 
accounting for zenith angles and offsets from the observations.

\begin{figure}[ht!]
\begin{minipage}[c][4.6cm][c]{0.3\textwidth}
\centering
\includegraphics[height=4.5cm,angle=0]{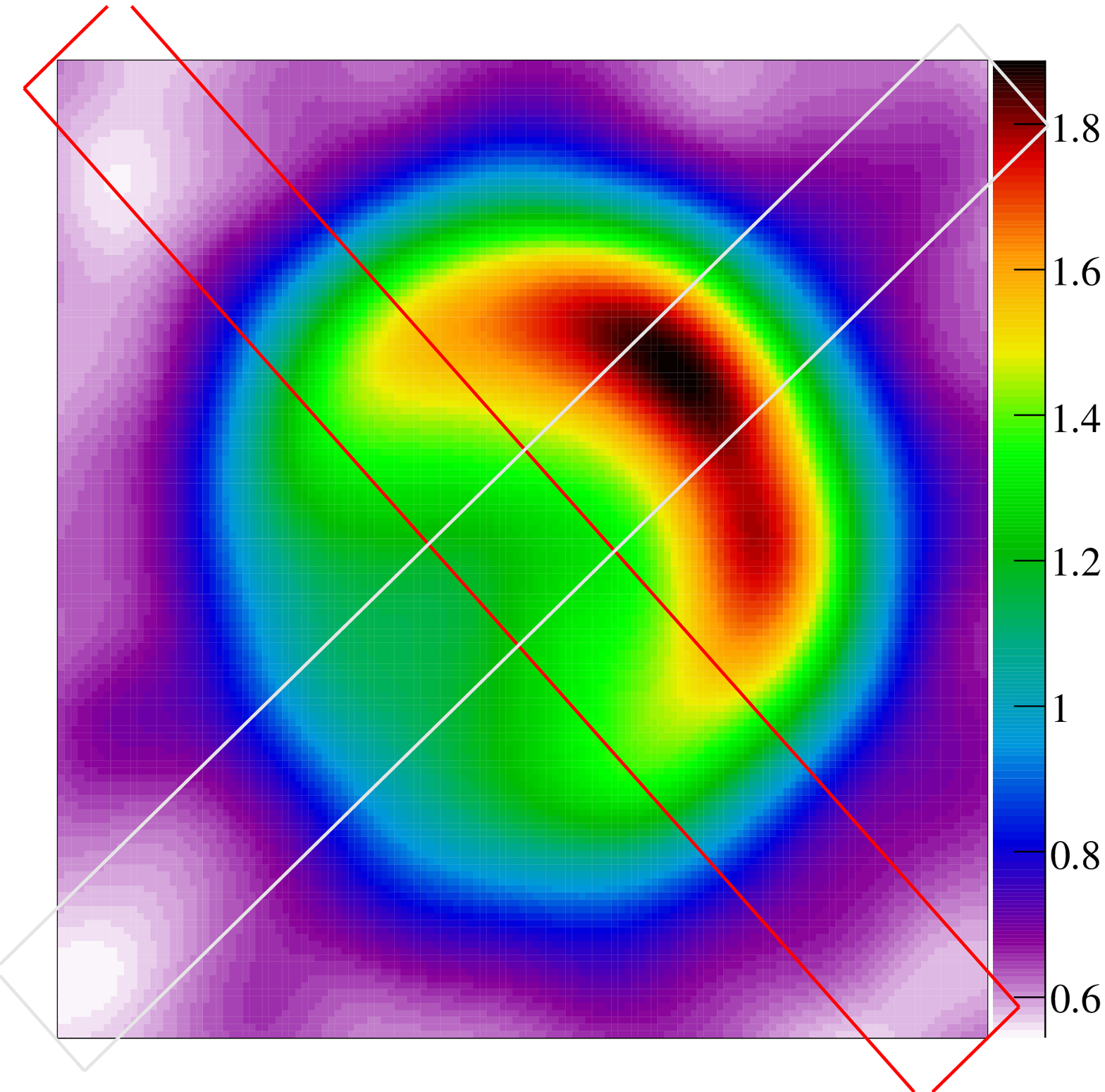}
\end{minipage}
\begin{minipage}[c][4.6cm][c]{0.3\textwidth}
\centering
\includegraphics[height=4.5cm,angle=0]{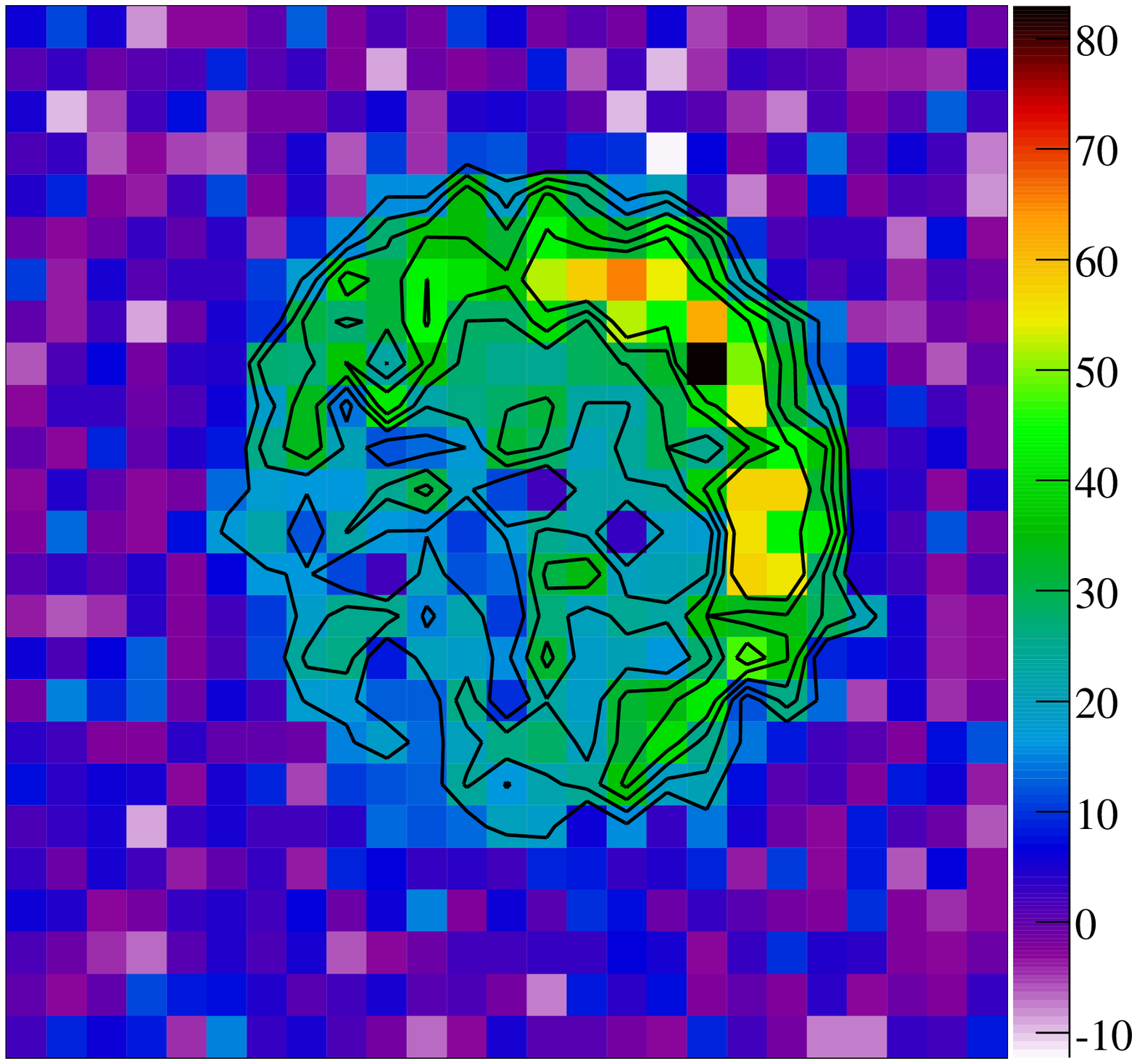}
\end{minipage}
\begin{minipage}[c][4.6cm][c]{0.3\textwidth}
\centering
\includegraphics[height=4.5cm,angle=0]{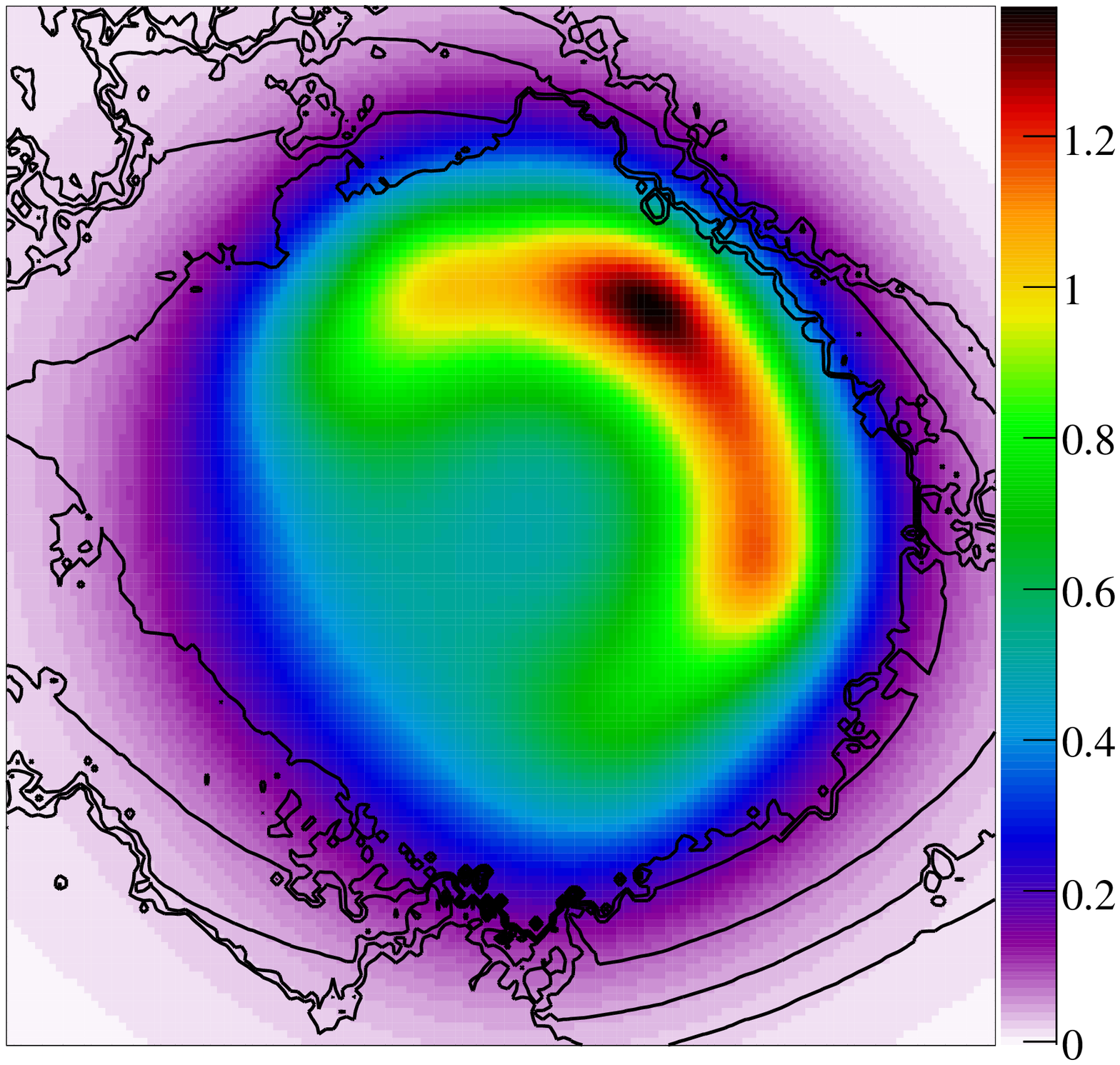}
\end{minipage}
\caption{Applying different methods to the same count map. From {\it left} to {\it right}, 
fixed-scale smoothing, re-binning, ASMOOTH. The contours are for
significance, from 3 to 7 $\sigma$ or 5 to 9$\sigma$ for ASMOOTH). \label{figure:methods1} }
\end{figure}

\begin{figure}[h!]
  \includegraphics[width=16cm,angle=0]{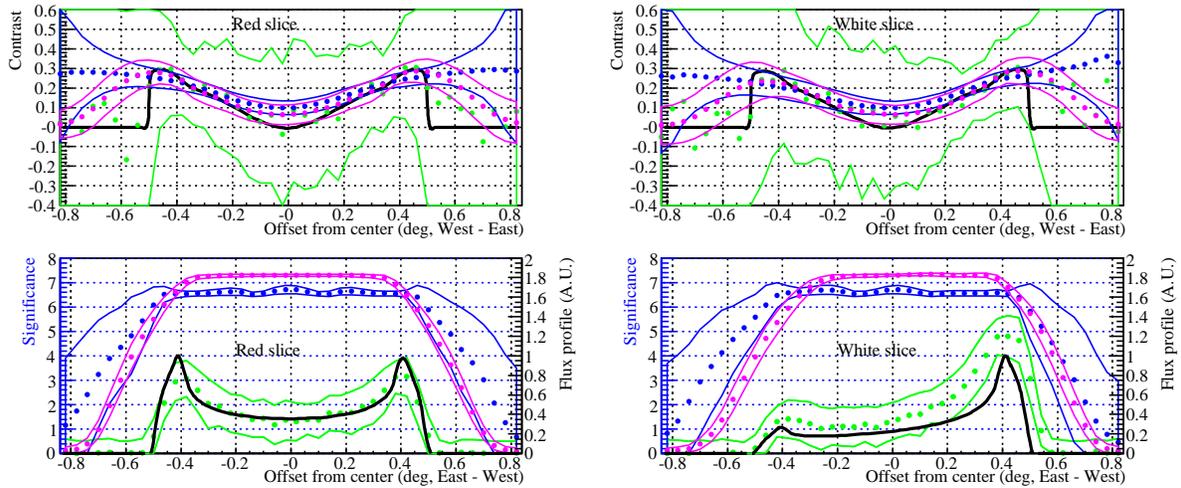}
  \caption{Slices for model I,
    with the geometry of Figure~\ref{figure:methods1}{\it left}. In black is shown the model.
    Green, blue and pink are respectively the fixed binning (0.04$^{\circ}$ integration radius), WVT and
    ASMOOTH algorithms, with 1 $\sigma$ error bands. }
  \label{figure:slicesL}
\end{figure}

The acceptance profiles are built for the same energy bands from regions in
the sky devoid of $\gamma$-ray source.
For each position on the map, the spectral power-law index is varied
until the hardness ratio agrees with the data. The flux is then scaled to the excess.
\newline
{\bf Disclaimer}\\
Several important issues are not investigated here, notably the 
variation of the PSF with energy (smaller at higher energy) and its effect on the actual
spatial resolution, nor the inclusion of trial factors in the significance.

\begin{figure}[ht!]
  \includegraphics[width=16cm,angle=0]{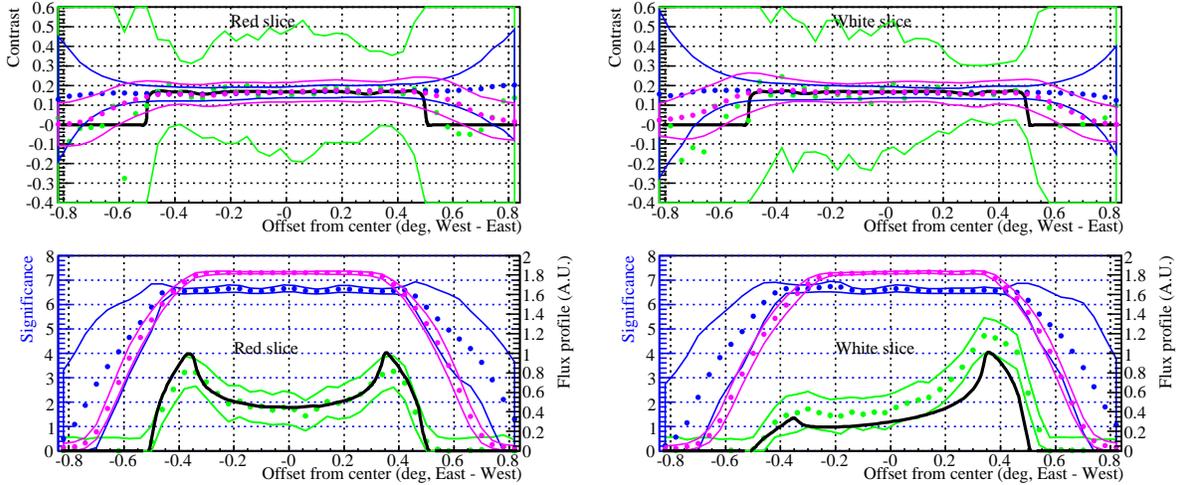}
  \caption{Slices for model II, following the conventions
    of Fig.\ref{figure:slicesL}. }
  \label{figure:slicesH}
\end{figure}

\section{Applications}
{\bf Simulating a shell-type supernova remnant}\\
\label{section:shell}
RX~J1713-3947 is a large shell-type SNR with bright non-thermal X-ray
emission~\cite{c:koyama97,c:slane99}. 
Recent XMM~\cite{c:cassam04} and Nanten~\cite{c:moriguchi05} observations 
suggest interactions with molecular clouds, at $\sim$ 1.3 kpc.
TeV spectro-imaging was obtained by HESS~\cite{c:aharonian06,c:aharonian07},
showing good correlation with the X-ray morphology. The TeV spectral index seems 
constant across the remnant, at odds with X-rays~\cite{c:cassam04}.
The nature of the TeV emission is still 
debated~\cite{c:uchiyama07,c:butt08,c:ellison08,c:plaga08}. 

We investigate the visibility by HESS of two qualitative models: 
one (model I) with a spectral photon index decreasing from the shock front (-2) 
to the center (-2.4), the other (model II) with a constant index (-2.1). 
The flux distribution follows a shell-type profile (spherical geometry) 
with a gradient from South-East to North-West.
50 simulations, mimicking the current HESS statistics in the energy
bands 0.45 - 1 TeV and 1.2 - 30 TeV , were produced to estimate the fluctuations.\\
The contrast profiles, Fig.~\ref{figure:slicesL} and~\ref{figure:slicesH}, are properly recovered
by the three methods. The significance values are also as expected: varying with the flux
level for the fixed binning and stable near the requested level
everywhere within the emission region for the other methods.
For the same reason, small scale details are lost, as shown for model I
by the biased contrast given by ASMOOTH and WVT at the center of the shell.
This suggests that fixed binning may have more potential for the discovery of sources or emission features,
to be confirmed by additional observation.\\
Overall, from the errors on the contrast, the variation for model I is only a 2-3 $\sigma$ effect on average.
Discriminating between the two models included here would probably be
difficult on a real source.

\begin{figure}[ht!]
  \includegraphics[width=15cm,angle=0]{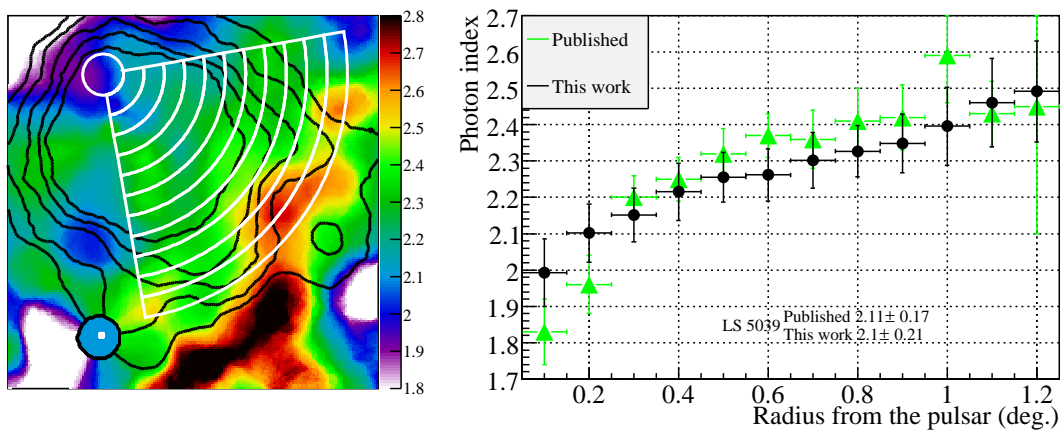}
  \caption{Photon index from the WVT maps. The significance contours (black) in the map 
    are for 3 to 7 $\sigma$). The values in the histogram correspond to the
    concentric slices (white) in the map, centered on the pulsar. \label{figure:gamma} }
\end{figure}

\begin{figure}[h!]
  \includegraphics[width=15cm,angle=0]{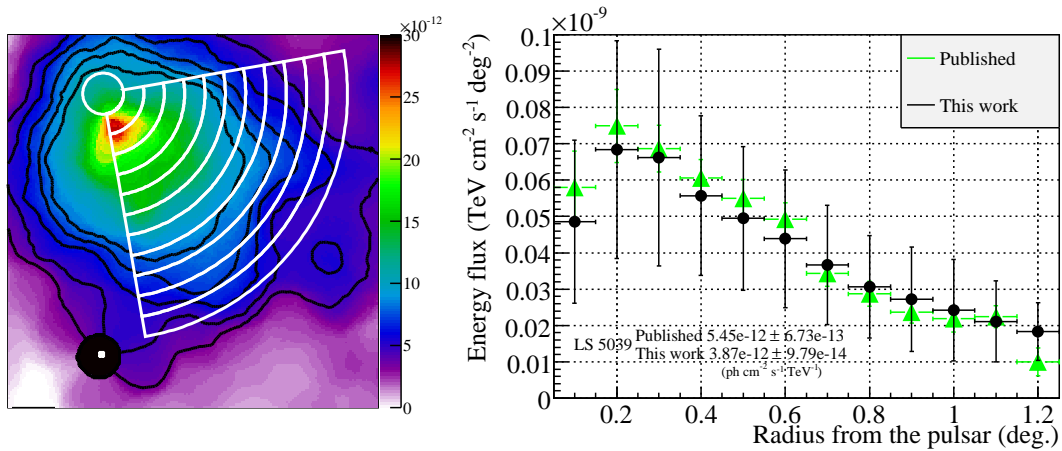}
  \caption{Photon flux density from the WVT maps. The map is scaled in ph.~cm$^{-2}$~s$^{-1}$~deg$^{-2}$.
    The contours and histogram are defined similarly to Fig.~\ref{figure:gamma}. \label{figure:flux} }
\end{figure}

{\bf Pulsar wind nebula HESS J1825-137}\\
\label{section:pwn}
HESS~J1825-137, detected in the HESS Galactic Survey~\cite{pwn:aharonian05},
is likely associated with the X-ray PWN G18.0-0.7 and the very
energetic (3$\times$10$^{36}$ erg/s) pulsar PSR J1826-1334~\cite{atnf}. 
The VHE emission is offset to the South / South-West of the pulsar.
$\gamma$-ray spectral indexes in the PWN~\cite{pwn:aharonian06}
soften with increasing distance to the pulsar.\\
From the energy bands 0.27 - 0.9 TeV and 1.1 - 30 TeV, 
we produced maps of the photon index, Fig.~\ref{figure:gamma},
and the flux density, Fig.~\ref{figure:flux}. The estimation of errors are preliminary.\\
Our results on the PWN are compatible with the HESS publications: the evolution of the photon 
index is clearly seen, with a similar range of values, and so is the flux normalization profile.
The superposed significance contours indicate that the external part of the profile is not reliable.\\
The neighbouring binary LS 5039, a point source for HESS, was treated as 
a fixed bin of radius 0.1~$^{\circ}$. The integrated flux and photon index are here again 
very close to the published~\cite{ls5039} phase-averaged numbers.

\section{Conclusion}

The methods discussed here give results compatible with HESS publication.
They may help exploring the morphology of extended VHE sources,
although as illustrated in a simulation, the uncertainties might prove too large for detailed studies.


\begin{theacknowledgments}
  We wish to thank the HESS collaboration for allowing us to use its data in this work.
\end{theacknowledgments}


\end{document}